\documentclass[aps,prb,floats,twocolumn]{revtex4}
\usepackage{epsfig}
\usepackage{amsmath}

\begin{document}         
\title{Curie temperature of Kondo lattice films with finite itinerant charge carrier density} 
\author{J. Kienert and W.~Nolting}
\affiliation{Festk{\"o}rpertheorie, Institut f{\"u}r Physik, Humboldt-Universit{\"a}t zu Berlin, 
  Newtonstr. 15, 12489 Berlin, Germany}
\begin{abstract}
We present a model study of ferromagnetic films consisting 
of free Bloch electrons coupled to localized moments (Kondo lattice films). 
By mapping the local interaction onto an effective
Heisenberg Hamiltonian we obtain temperature and carrier density dependent exchange
integrals mediating the interaction between local moments via the conduction electrons. 
The non-perturbative approach recovers analytically the
weak-coupling RKKY interaction and yields convincing numerical results in the strong 
coupling (double exchange) regime. The Curie temperature is calculated for various coupling 
strengths, band fillings, and numbers of layers. The results are compared with total energy 
calculations. We discuss the influence of charge transfer 
between film layers and of anisotropy on the Curie temperature. The model we investigate
is considered relevant for the understanding of the basic magnetic
properties of manganites, diluted magnetic semiconductors, and rare
earth substances as, e.g., Gadolinium.

\end{abstract}
\maketitle
\section{Introduction}
The Kondo lattice model (KLM) is one of the prototype models in solid state physics 
whenever the coupling of itinerant charge carrier spins with immobile localized 
magnetic moments has to be considered. There are various quite different classes 
of materials for which the KLM is used to describe the electronic and magnetic properties 
at least in principle. 

In its simplest single-band version the KLM Hamiltonian reads
\begin{equation}
  \label{Hklm}
  H= \sum_{ij
    \sigma}(T_{ij}-\mu\delta_{ij})c_{i\sigma}^{\dag}c_{j\sigma}-J\sum_{i}{\bf S}_{i}\cdot {\boldsymbol \sigma}_{i}\;. 
\end{equation}
The first term describes non-interacting electrons ($c_{i\sigma}^{(\dag)}$ annihilates (creates) an electron 
of spin $\sigma=\uparrow,\downarrow$ at lattice site $i$) with hopping integrals 
$T_{ij}$. ${\boldsymbol \sigma}_i$ is the electron spin and ${\bf S}_i$ the localized 
spin operator, and both are exchange coupled by $J$.

Depending on the sign of $J$ one is confronted with quite different physics.
Numerous studies have been carried to understand the physics of 
heavy fermion systems. Here the coupling between electrons and localized spins favors an 
antiparallel configuration ($J<0$) leading to a competition between RKKY interaction and spin 
screening effects.

In this work we want to focus on a favored parallel spin alignment ($J>0$, ferromagnetic Kondo lattice 
model, FKLM). The FKLM was introduced to model the magnetism in some manganese compounds ("manganites"). 
In the framework of a simple two-site model this kind of interaction was found to result 
from the so-called double exchange lending its name to the lattice version 
in the strong coupling regime.\cite{Zen51,AH55} 

In the manganites the 5 Mn d-shells are split by the crystal field into three degenerate
$t_{2g}$-orbitals forming localized spins of $S=\frac{3}{2}$ which
interact via Hund's rule with itinerant electrons stemming from the
remaining two degenerate $e_{g}$-orbitals. Due to the Jahn-Teller effect and the strong 
Coulomb interaction among the conduction electrons the manganites exhibit a complex phase 
diagram, the most prominent phenomenon of which is colossal magnetoresistance.\cite{Dag03,Ram97} 
Although for a more quantitative description of manganite materials additional ingredients like orbital 
physics, electron-polaron interaction, and a strong Hubbard-like Coulomb interaction are needed, numerous 
research work on the subject has been carried out using the simple one-orbital model (\ref{Hklm}).\cite{Dag03} 

More recently attention was directed to the diluted magnetic semiconductors (DMS) 
that are thought to have the potential for promising technical applications 
in microelectronics.\cite{Ohno99,ZFS04} Substituting transition metal 
impurities like Mn in a semiconducting host (often III/V like GaAs) localized magnetic moments 
are introduced with the consequence of ferromagnetic ordering of the randomly distributed maganese 
cation spins that interact via their (ferromagnetic, intermediate) coupling to valence and impurity 
band holes.

Another important field of application of the FKLM is the theoretical description of rare earth 
substances like Eu and Gd and their compounds. Here it initially served to explain the famous 
red-shift of the absorption edge of the optical $4f-5d$ transition in the ferromagnetic 
semiconductor EuO.\cite{Wach64} Combined LDA/many-body calculations based on the FKLM yielded 
realistic values of the Curie temperature for Gd\cite{SEN04} and predicted surface states in 
EuO-films\cite{ScN01}. Doping Gd into EuO makes for an attractive means to tune the charge carrier density 
and thus the carrier-induced coupling among the magnetic ions.\cite{Ott06} Having the same benefit, 
Gd-doped GaN was reported to yield high $T_C$ ferromagnetism above 300 K.\cite{Ploog05} 

In all of the above-mentioned substances, especially regarding possible technical applications 
in the future, it is important to understand and make use of the special physical properties due to 
reduced dimensionality. Several recent works focussed on the role of magnetic anisotropy 
effects in thin films of manganites\cite{Sid06} or of DMS films like (Ga,Mn)As\cite{Wang05}. Gd films 
are investigated due to their potential employment in spintronics as well as motivated by the "evergreen" issue 
of a possibly enhanced surface Curie temperature.\cite{Wand98,SPF00,Nick04} It is furthermore a well-known 
fact that the Curie temperature strongly depends on the film thickness. Due to the broad relevance of the FKLM 
in current research on magnetism and electronic correlations and due to the importance of film structures it is 
worthwhile investigating the dependence of the ferromagnetic transition temperature on the model 
parameters.

From the theoretical point of view our treatment of FKLM films that is divided up into two parts. 
First we solve the problem for the electronic self-energy within an equation of motion approach 
that is based on a decoupling scheme 
fulfilling non-trivial limiting cases. In the second step we map the interaction onto an effective 
localized spin Hamiltonian by integrating out the electronic degrees of freedom. This results in 
effective exchange integrals which incorporate the interaction of the localized moments via the conduction 
electrons beyond conventional RKKY theory. They depend on temperature and carrier density. In sum we end up 
with a self-consistent theory which does not require the assumption of classical core spins and which is 
employable at all temperatures. 

Results on the magnetization and critical temperature of ferromagnetic KLM model films with finite band 
occupation have already been reported in Ref. \onlinecite{KN04}. However, in the present work we make use 
of a more sophisticated approach for the fermionic subsystem and carry out a wider and considerably more 
detailed investigation of the Curie temperature. A similar system is also treated in Ref. \onlinecite{Urb94} 
based on a method guided by the same spirit. However, the theory of the electronic subsystem there is on a 
mean-field level only and just nearest neighbor Heisenberg interaction is considered, in spite of the underlying 
RKKY-like exchange being long ranged. What's more the effective exchange integrals derived in 
Ref. \onlinecite{Urb94} in terms of the susceptibility of free electrons are proportional to $J^2$ and thus 
fail to give the appropriate strong-coupling behavior where the Curie temperature is independent of $J$. As 
far as the numerical evaluation is concerned the thickness dependence of the Curie temperature is analysed.

This paper is organized as follows. In the next section we briefly summarize our method to solve 
the KLM for the electronic self-energy. Then we derive an effective Heisenberg Hamiltonian which 
results from integrating out the fermionic degrees of freedom. In the numerical results section 
we discuss the temperature dependent quasiparticle excitation spectrum and charge transfer before we 
present an extensive analysis of the Curie temperature and its dependence on the various model paramters 
like, e.g., the magnetic anisotropy strength.

\section{Electronic self-energy}

In second-quantized notation and introducing Greek indices to number the layers the 
Hamiltonian (\ref{Hklm}) becomes
\begin{eqnarray}
\label{Hs}
H&=&H_{0} + H_J\\
 &=&\sum_{ij\sigma\atop\alpha\beta}(T_{ij}^{\alpha\beta}-\mu\delta_{ij}^{\alpha\beta})c_{i\alpha\sigma}^{\dag}c_{j\beta\sigma}\\
\label{Hsf}
 &-&\frac{1}{2}J\sum_{i\alpha\sigma}\left(z_{\sigma}S_{i\alpha}^{z}n_{i\alpha\sigma}+S_{i\alpha}^{-\sigma}c_{i\alpha\sigma}^{\dag}c_{i\alpha-\sigma}\right)\;.
\end{eqnarray}
$S_{i\alpha}^{+,-}=S_{i\alpha}^{x}\pm iS_{i\alpha}^{y}$, $\uparrow(\downarrow)=+~(-)$, $z_{\uparrow}=1,~z_{\downarrow}=-1,$ and
$S_{i\alpha}^{x,y,z}$ are the cartesian components of the localized spin operator ${\bf S}_{i\alpha}$. 
$n_{i\alpha\sigma}$ is the particle density operator.

In the following we apply an equation of motion (EOM) approach based on a decoupling scheme already 
applied to semiconducting Kondo lattice films (EuO\cite{ScN01} and EuS\cite{MN04}) 
and to bulk Gd\cite{SEN04} in combined LDA and many-body calculations
(moment conserving decoupling approximation, MCDA). For details of the 
method we refer the reader to corresponding model studies.\cite{SN99,NRM97,SN02} After writing down the 
EOM of the one-particle Green function $G^{\alpha \beta}_{ij\sigma}(E)=\langle\langle
c_{i\alpha\sigma};c^{\dag}_{j\beta\sigma}\rangle\rangle$ and the generated higher Green functions 
a decoupling is performed that is guided by some non-trivial cases, as, e.g., the ferromagnetic semiconductor 
at T=0. In the ${\bf k}$-space defined within the film plane and using matrix notation one can write:
\begin{eqnarray}
\label{G}
{\bf G}_{{\bf k}\sigma}(E)=\left[E{\bf I}-{\mathbf \epsilon}({\bf k})-{\mathbf \Sigma}_{{\bf
      k}\sigma}(E)\right]^{-1}\;.
\end{eqnarray}
${\bf \epsilon}({\bf k})$ is the hopping matrix. A local
approximation of the self-energy
$\Sigma^{\alpha\beta}_{{ij}\sigma}(E)\rightarrow
\delta_{ij}\delta_{\alpha\beta}\Sigma_{\alpha\sigma}(E)$ is performed. This
corresponds to neglecting magnon energies in the three-dimensional case\cite{roldiss00} and is justified by the
fact that these energies are usually orders of magnitude smaller than
the exchange coupling $J$ and the electron bandwidth $W$.

From (\ref{G}) one immediately obtains the one-particle local density of
states (LDOS):
\begin{eqnarray}
\label{ldos}
\rho_{\alpha\sigma}(E)=-\frac{1}{\pi\hbar}{\rm Im}G^{\alpha\alpha}_{ii\sigma}(E+i0^{+}-\mu)\;.
\end{eqnarray}

It should be mentioned that in general the MCDA approach does not capture Kondo scaling and especially violates 
the Luttinger theorem ${\rm Im}\Sigma_{\sigma}(E=\epsilon_{F})\rightarrow0$.\cite{BMG05} Although this might well be due to our approximative scheme
questions if and in which parameter regimes the ferromagnetic Kondo lattice model is a Fermi liquid still remain open. Nevertheless we would 
consider it worthwhile to investigate the low temperature Kondo physics within the framework of our theory. In the present work, however, 
we rather want to focus on the magnetic transition temperature.  

The self-energy matrix ${\mathbf \Sigma}_{\sigma}(E)$ depends on various
expectation values of pure fermionic, mixed fermionic-spin, and pure
localized spin character:
\begin{eqnarray}
{{\Sigma}^{\alpha}_{\sigma}}=F(\langle n_{\alpha\sigma}\rangle,
\langle
S_{\alpha}^{-\sigma}c_{\alpha\sigma}^{\dag}c_{\alpha-\sigma}\rangle,~\langle
  S_{\alpha}^{z}n_{\alpha\sigma}\rangle,\\\nonumber
\langle{S^{z}_{\alpha}}\rangle,~\langle{(S^{z}_{\alpha})^2}\rangle,~\langle{(S^{z}_{\alpha})^3}\rangle,~\langle{S^{+}_{\alpha}S^{-}_{\alpha}}\rangle)\;.
\end{eqnarray}

The first two types can be calculated self-consistently within the MCDA using the 
corresponding Green functions and the spectral theorem\cite{Nol7}.
The localized spin correlation functions like, to lowest order, 
the layer-dependent magnetization  $\langle S_{\alpha}^z \rangle$ need 
further consideration. In order to evaluate these we map the interaction term (\ref{Hsf}) 
onto an effective coupling between the localized spins, which is carried out in the next section. 

\section{Effective localized spin Hamiltonian}

It is well known that using perturbation theory (\ref{Hsf}) leads to the so-called RKKY
interaction.\cite{RKKY,NolQdM} This second order interaction between localized
spins is long ranged and yields an oscillatory behavior of the exchange
coupling which is mediated by uncorrelated electrons. It appears highly
desirable to have such a mapping without being restricted to the weak
coupling regime. This requires finding an effective Hamiltonian
\begin{eqnarray}
\label{Hklm_av}
H^{eff}_{J}&=&\langle H_{J} \rangle^{(c)}\\
\nonumber
&=&-\frac{J}{2N}\sum\limits_{i\alpha\sigma\sigma^{\prime}\atop {\bf kq} }e^{-i{\bf
        q}{\bf R}_{i}^{\alpha}}\left({\bf S}_{i\alpha}\cdot{\boldsymbol \sigma}\right)_{\sigma\sigma^{\prime}}\langle c^{\dag}_{{\bf k}+{\bf
        q}\alpha\sigma}c_{{\bf k}\alpha\sigma^{\prime}}\rangle^{(c)},
\end{eqnarray}
where $N$ is the number of lattice sites in one layer and the superscript $c$ formally indicates that the 
averaging is performed in the fermionic subspace only. The method we present in this section has
been applied before to bulk Kondo lattice systems.\cite{SN02,SEN04} We
generalize it into a matrix formulation which allows us to consider
film structures, but is also relevant whenever sublattice
decompositions are necessary, for instance when dealing with
antiferromagnetic configurations.
 
In order to obtain the expectation value in (\ref{Hklm_av}) we
introduce the modified Green function   
\begin{eqnarray}
\label{Gmod}
{\hat G}_{{\bf k},{\bf k}+{\bf
      q}\alpha\beta}^{\sigma^{\prime}\sigma}(E)=\langle\langle c_{{\bf
      k}\alpha\sigma^{\prime}};c^{\dag}_{{\bf k}+{\bf q}\beta\sigma}\rangle\rangle^{(c)}\;.
\end{eqnarray}
The EOM for ${\hat G}$ reads  
\begin{eqnarray}
\label{EOM}
    \sum_{\gamma}\left(E\delta_{\gamma\alpha}-\epsilon^{\alpha\gamma}_{\bf k}\right){\hat G}_{{\bf k},{\bf k}+{\bf
        q}\gamma\beta}^{\sigma^{\prime}\sigma}(E) = \delta_{{\bf
        q}0}\delta_{\sigma\sigma^{\prime}}\delta_{\alpha\beta}\\
     -\frac{J}{2N}\sum_{i{\bf p}\sigma^{\prime\prime}}e^{i({\bf
        k}-{\bf p}){\bf R}_{i}^{\alpha}}\left({\bf S}_{i\alpha}\cdot{\mathbf
        \sigma}\right)_{\sigma^{\prime}\sigma^{\prime\prime}}{\hat G}_{{\bf p},{\bf k}+{\bf
        q}\alpha\beta}^{\sigma^{\prime\prime}\sigma}(E)\;.
\end{eqnarray}
In principle, this equation can be solved iteratively. Applying the
spectral theorem then yields the expectation value in
(\ref{Hklm_av}). This correlation function is an operator since
we are working in the fermionic subspace, resulting in an effective
interaction among the localized spins ${\bf S}_i$ only.

We have to find a manageable approximation for ${\hat G}$ on the right
hand side of (\ref{EOM}). It can easily be shown that in the
non-interacting limit,
\begin{eqnarray}
\label{crkky}
  {\hat G}_{{\bf p},{\bf k}+{\bf
      q}\alpha\beta}^{\sigma^{\prime\prime}\sigma}(E) \rightarrow \delta_{\sigma^{\prime\prime}\sigma}\delta_{{\bf
      p},{\bf k}+{\bf q}}G_{{\bf k}+{\bf
      q}\alpha\beta}^{(0)}(E)\;,
\end{eqnarray}
conventional RKKY interaction is reproduced. In order to include spin scattering effectively we
dress the free propagator in (\ref{crkky}) and replace it by the full
propagator $G_{{\bf k}+{\bf q}\alpha\beta\sigma}(E)$. Within this
modified RKKY approximation (MRKKY) the exchange integrals in the effective
Hamiltonian
\begin{eqnarray}
\label{HJeff}
H_{J}^{eff}=-\sum\limits_{ij\alpha\beta}{\hat
      J}_{ij}^{\alpha\beta}{\bf S}_{i\alpha}{\bf
      S}_{j\beta} 
\end{eqnarray}
are given by
\begin{eqnarray}
\nonumber
\label{Jeff}
 {\hat J}^{\alpha\beta}({\bf q})&=&\frac{J^2}{4\pi N}\sum\limits_{{\bf
     k}\sigma}{\rm Im}
    \int\limits_{-\infty}^{+\infty}dE f(E)G_{{\bf
        k}\sigma}^{(0)\alpha\beta}(E)G_{{\bf k}+{\bf
        q}\sigma}^{\alpha\beta}(E)\\
{}
\end{eqnarray}
where $f$ is the Fermi function and the sum is over the first Brillouin
zone.

Within the Tyablikov approximation\cite{Tya69,NolQdM} the imaginary part of the transversal 
layer-diagonal spin Green function (spectral density) can be written as  
\begin{eqnarray}
-\frac{1}{\pi}{\rm Im}\langle\langle S^{+}_{{\bf k}\alpha}; S^{-}_{(-{\bf
    k})\alpha}\rangle\rangle
=2\langle{S_{\alpha}^z}\rangle\sum_{\gamma}\eta^{\alpha\gamma}_{\bf
  k}\delta\left(E-E_{\gamma}({\bf k})\right) 
\end{eqnarray}
where the spectral weights $\eta^{\alpha\gamma}_{\bf k}$ and energy poles
$E_{\gamma}({\bf k})$ have to be evaluated numerically. For a monolayer
we can drop all Greek indices, $\eta_{\bf k}=1$, and we get the
well-known spin wave energies 
\begin{eqnarray}
\label{E_k}
E({\bf k})=2\langle S_{z}\rangle({\hat J}({\bf 0})-{\hat J}({\bf k}))\;.
\end{eqnarray}

Applying the Callen method\cite{Cal63,NolQdM} to superlattices the layer-dependent 
magnetization,
\begin{eqnarray}
\label{Sz}
\langle{S_{\alpha}^z}\rangle
=\frac{\displaystyle(1+\varphi_\alpha+S)\varphi^{2S+1}_{\alpha}+(S-\varphi_{\alpha})(1+\varphi_{\alpha})^{2S+1}}{\displaystyle
 (1+\varphi_{\alpha})^{2S+1}-\varphi^{2S+1}_{\alpha}}
\end{eqnarray}
and other higher order spin correlation functions can be obtained using the Bose-like distribution function
\begin{eqnarray}
\label{phi}
\varphi_{\alpha}=\frac{1}{N}\sum_{{\bf
  k},\gamma}\frac{\eta^{\alpha\gamma}_{\bf k}}{e^{\beta
    E_{\gamma}({\bf k})}-1}\;.
\end{eqnarray}
(\ref{G}),(\ref{Jeff}), and (\ref{Sz}) represent a self-consistent
system of equations that can be solved for the one-particle Green
function matrix ${\bf G}_{{\bf k}\sigma}(E)$ and the magnetization
$\langle{S_{\alpha}^z}\rangle$.

Before proceeding to the numerical evaluation we have to reconsider our
effective Hamiltonian (\ref{HJeff}). It is known that for
low-dimensional systems anisotropies can become very important and even
a necessary condition for magnetic ordering at finite
temperature.\cite{MW66,GN01} 
We therefore include a single-ion anisotropy term
\begin{eqnarray}
\label{H_A}
H_{A}=-K^{\alpha}_2 \sum_{i\alpha}(S^{z}_{i\alpha})^2\;.
\end{eqnarray}
The physical background of this magnetic anisotropy is spin-orbit
coupling, which usually is some order of magnitudes smaller than the
exchange coupling between localized spins. A positive $K^{\alpha}_2$ favors an 
out-of-plane easy axis, i.e. perpendicular to the film plane.

We treat this term in the Anderson-Callen approximation\cite{AC64} and decouple the higher
Green function generated by (\ref{H_A}) in the following manner:
\begin{eqnarray}
\label{GF_aniso}
\langle\langle\left[S_{i\alpha}^{+},H_{A}\right]_{-};S_{j\beta}^{-}\rangle\rangle&=&\langle\langle
S_{i\alpha}^{+}S_{i\alpha}^{z}+S_{i\alpha}^{z}S_{i\alpha}^{+};S_{j\beta}^{-}\rangle\rangle\nonumber\\ 
&\approx& \Phi_{\alpha}\langle\langle S_{i\alpha}^{+};S_{j\beta}^{-}\rangle\rangle\;,
\end{eqnarray}
\begin{equation}
\Phi_{\alpha}=2\langle S^{z}_{\alpha}\rangle \left(1-\frac{1}{2S^2}\left(S(S+1)-\langle
    (S^{z}_{\alpha})^2\rangle\right)\right)\;.
\end{equation}
We can drop the site index $i$ due to lateral translational invariance.
The single-ion anisotropy acts as an effective field $\Phi_{\alpha}$ 
coupled to $S^{z}_{\alpha}$.
Note that this decoupling is valid only if the magnetization is parallel to
the $z$-direction. In order to ensure this a rotation of the coordinate
system might be required before the decoupling, for instance when an arbitrary oriented 
magnetic field is present.\cite{SKN05}

For bulk or a monolayer ($\alpha=1$) it is possible 
to derive a simple formula for the Curie temperature $T_C$. Expanding (\ref{Sz}) and 
(\ref{phi}) in the vicinity of $T_C$ one easily finds
\begin{eqnarray}
\label{Tc}
\frac{3k_{B}T_{C}}{S(S+1)N} = \left(\sum\limits_{\bf
      q}\frac{1}{\left(K_{2}\gamma +2({\hat
      J}({\bf 0})-{\hat J}({\bf q}))\right)}\right)_{T=T_{C}}^{-1}
\end{eqnarray}
where 
\begin{eqnarray}
\gamma = \lim_{T\to T_{C}}\frac{\Phi(T)}{\langle
  S^{z}\rangle}=\frac{2(2S-1)}{3S}
\end{eqnarray}
is a constant and depends on the specific decoupling of the anisotropy 
Green function (\ref{GF_aniso}).\cite{remark1}

For $\alpha>1$ the spin wave energies are no longer known in analytical form 
as in (\ref{E_k}). One can then, however, use (\ref{Sz}) to determine 
the critical temperature.

\section{Numerical results}

All numerical results have been obtained for simple 
cubic (sc) (100) films. The nearest-neighbor hopping integral $t$ in the tight binding approximation 
was chosen according to a bulk bandwidth $W=1~{\rm eV}$, i.e. $t=-0.083~{\rm eV}$. The value of 
the exchange coupling is assumed to be homogeneous and isotropic for all sites and 
layers of the films, as already implied by (\ref{Hsf}). Furthermore for the 
anisotropy constant we set $K_{2}^{\alpha}=K_2$ for all layers.
The magnitude of the localized spin is $S=7/2$. To keep notation simple we write $n_{(\alpha,\sigma)}$ 
for the respective expectation values $\langle n_{(\alpha,\sigma)} \rangle$. $n$ denotes the average 
electron density.

\begin{figure}[t]
\epsfig{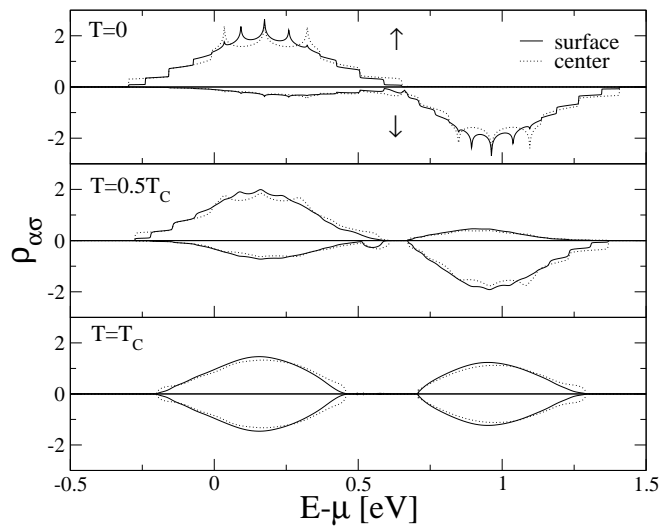}
\caption{One-particle density of states of a 5-layer film at different
  temperatures. Parameters: $n=0.2,~J=0.2~{\rm eV},~S=7/2$. Critical temperature
  $T_C=174~$K. Charge neutral calculation.}
\label{qdos_Tscan}
\end{figure} 

\subsection{Electronic properties}
\subsubsection{One-particle excitation spectrum}

We start with the discussion of the local one-particle density of
states. Fig. \ref{qdos_Tscan} shows the LDOS at different 
temperatures between $T=0$ and $T_C$. The Curie temperature has been
self-consistently computed.

The general picture for a layered system is the same as for the
bulk Kondo lattice model.\cite{SN02} At low temperature the spin-up
spectrum basically consists of the rigidly shifted non-interacting electron energy
band. At $T=0$ and in the insulating limit (band occupation $n=0$) this
is a rigorous result.\cite{Nol7} For the spin-down LDOS the situation is
different. Besides a band at higher energies corresponding to an
antiparallel coupling of conduction electrons and localized spins there is
a scattering part in the energy range of the $\uparrow$-band reflecting
the fact that a $\downarrow$-electron can flip its spin by creating 
a magnon in the localized spin subsystem. With increasing
temperature spin symmetry in the spectrum is gradually established. This is
accompanied by a reduction of the bandwidth due to reduced effective
hopping by spin scattering. For intermediate coupling strengths $J\approx
W$ as in Fig. \ref{qdos_Tscan} this leads to a temperature-induced
opening of a gap in the excitation spectrum. From the zero bandwidth
limit one can learn that the distance between the two correlated
bands roughly scales as $\sim J(S+\frac{1}{2})$.

Apart from these facts already known from bulk results one also
observes typical features of reduced dimensionality. The number
of van-Hove singularities which are most pronounced at low
temperatures is indicative of the number of layers and therefore for the
finiteness of the film system. A second important observation is the
reduced {\it effective} bandwidth of the surface compared to the center layer
LDOS. This can be directly traced back to the lower variance of the
surface LDOS: in the non-interacting limit $\Delta^{2}\rho_{0}=qt$ and
$q_{surface}/q_{bulk}=5/6$ for a sc(100) geometry. The effect is thus
not correlation induced and more pronounced for "open" film geometries. 
The band edges of the surface and center LDOS, however, are the same.

\subsubsection{Charge transfer}

The broken translational symmetry is furthermore responsible for the
occurence of charge transfer, i.e. the deviation $\delta^{\alpha} n$ of the
layer dependent band filling from the average occupation number: 
\begin{eqnarray}
\label{delta_charge}
\delta^{\alpha} n = n_{\alpha}-n =
\sum_{\sigma}(n_{\alpha\sigma}-\frac{1}{N_{L}}\sum_{\gamma}n_{\gamma\sigma})\;.
\end{eqnarray}
$N_{L}$ is the number of layers. Charge transfer is already present in the free 
electron film system and is due to the smaller effective bandwidth of the surface layer (occupation $n_{s}$). 
From this follows directly that in order to ensure thermodynamic equilibrium, i.e. a common
chemical potential $\mu$, one has $n_s<n$ below and $n_s>n$ above half-filling ($n=1$). 

\begin{figure}[t]
\epsfig{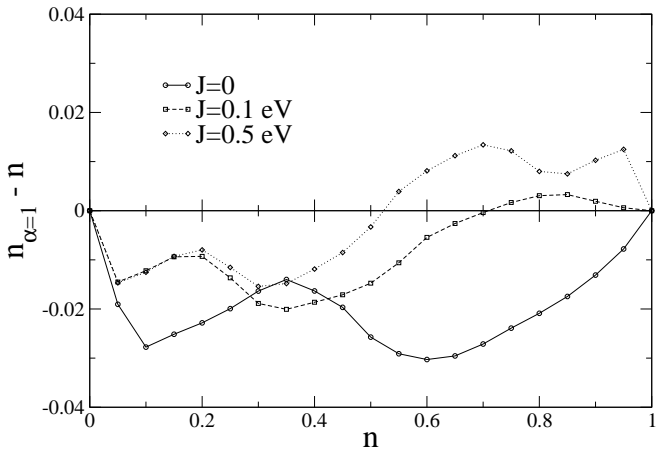}
\caption{Charge transfer in a 5-layer film at $T=0,~\langle S_{z}\rangle=S$
  (ferromagnetic saturation). Whereas the carrier density in the surface layer is always 
reduced in the non-interacting system there is an occupation above average for large enough $n$ in 
the correlated system.}
\label{charge_transfer_n}
\end{figure} 

Switching on the coupling of the electron spins to the localized spins the question
arises how this behavior changes. Fig. \ref{charge_transfer_n} shows the
difference of the surface and the average occupation number as a
function of $n$ at $T=0$ and ferromagnetic saturation. Whereas the electron density in the 
surface layer $n_{\alpha=1}$ is always below the average density $n$ in
the non-interacting limit ($J=0$) one also finds $\delta^{\alpha} n>0$ at
$n<1$ for $J>0$. The explanation follows the same reasoning as above for
the free system: a reduced or enhanced occupation in the surface layer
is linked to the chemical potential being in the lower or upper 
half of the LDOS, respectively. As there is a (mainly
$\downarrow$-) transfer of spectral weight to higher energies for 
finite $J$ the chemical potential crosses the maximum of the lower band
at values $n<1$. One finds the qualitatively same and quantitatively
similar behavior in the spin disordered phase where 
$\langle S^{\alpha}_{z}\rangle=0$.  

\begin{figure}[t]
\epsfig{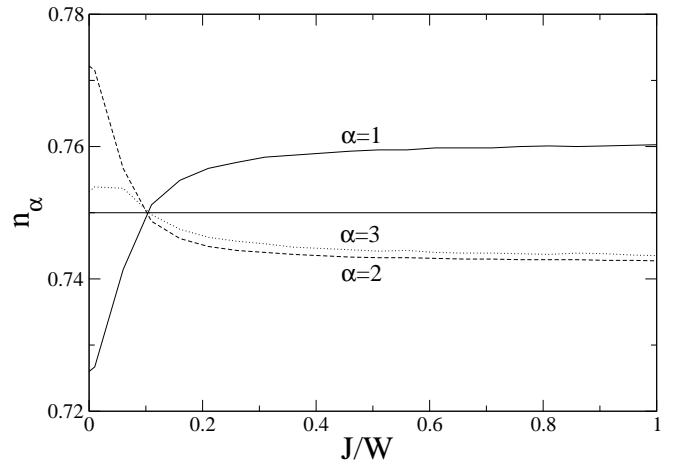}
\caption{Charge transfer in a 5-layer film at $T=0,~\langle
  S_{z}\rangle=0$. The horizontal line at $n=0.75$ indicates the total average occupation number. 
The choice of the energy scale refers to the free simple cubic bulk bandwidth $W=12t$.}
\label{charge_transfer_J}
\end{figure} 

A second point is the dependence of the charge transfer on the coupling
strength $J$. The interaction term (\ref{Hsf}) alone energetically favors single 
occupation of a lattice site for $n<1$. This can be easily seen in the zero bandwidth 
limit.\cite{Nol7,Nol78} It can thus be expected that a finite $J$ suppresses charge 
transfer. This is indeed the case as Fig. \ref{charge_transfer_J} shows where the electron density for 
the different layers of a 5-layer film is plotted as a function of the exchange coupling. The 
change of sign of $\delta^{\alpha} n$ is again caused by the same mechanism as pointed out above.
Given a band occupation of $n<1$ the chemical potential moves from the lower half of the Bloch 
bands ($J=0$) to the upper half of the lower quasiparticle subband ($J>0$).

We observe that even $J\rightarrow \infty$ does not lead to complete 
charge neutrality and that at large $J$ the higher occupation of the surface layer goes with
an almost homogeneous charge distribution in the inner layers of the film. 
It is also interesting to note that at $J\approx
0.1$ there is a quasi-homogenous charge distribution which coincides
with the change of sign of $\delta^{\alpha} n$.

One can summarize that the effect of charge transfer is highest at low
couplings $J$ and low average band occupations $n$, where it can amount
to almost 30\% (Fig. \ref{charge_transfer_n} at $n=0.1$). It appears to
be of minor importance in the strong coupling regime, especially at
higher charge carrier densities.

All the findings we have discussed in this section are qualitatively the same for other
film thicknesses. As already mentioned charge transfer effects are
more pronounced in "open" surface geometries with a stronger relative
reduction in the number of nearest neighbors of a surface atom. However,
the {\em order of magnitude} of the effect is the same. 

\begin{figure}[t]
\epsfig{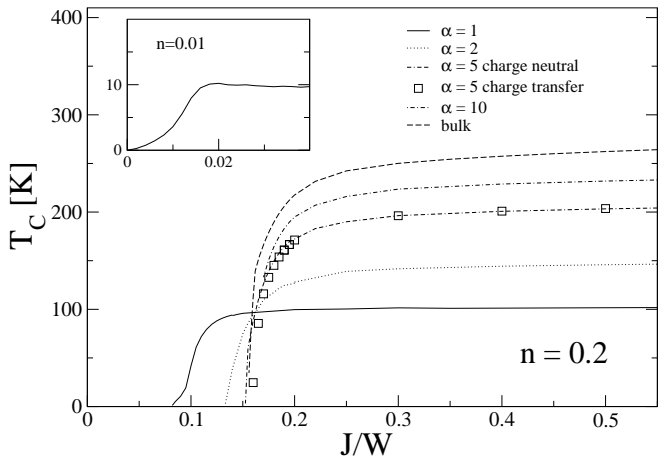}
\caption{Curie temperature for various numbers of layers. Parameters: $n=0.2, K_{2}=10^{-6}~{\rm eV}$. 
There is no significant influence of charge transfer on the critical temperature. 
Inset: Curie temperature of a monolayer for a low band occupation.}
\label{Tc_J_n0_2}
\end{figure} 
\begin{figure}[b]
\epsfig{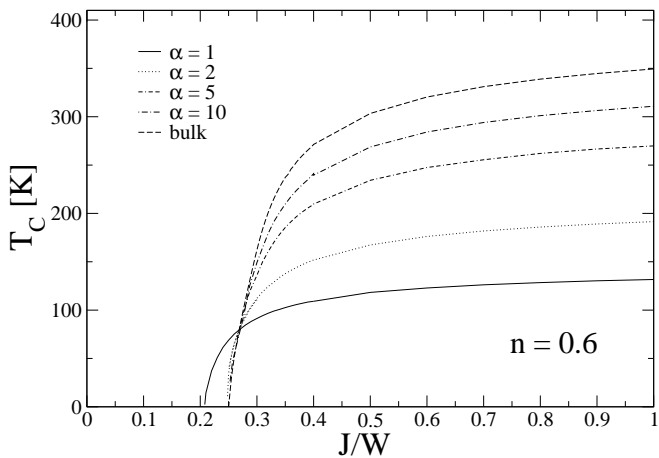}
\caption{Curie temperature for various numbers of layers. Parameters: $n=0.6, K_{2}=10^{-6}~{\rm eV}$.}
\label{Tc_J_n0_6}
\end{figure} 
\begin{figure}[t]
\epsfig{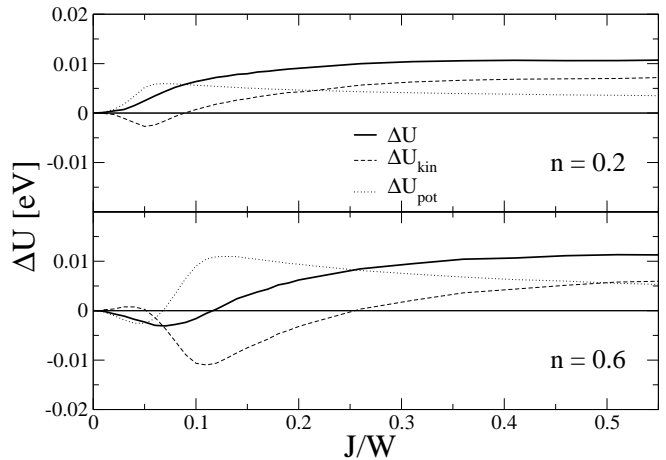}
\caption{Total, kinetic, and potential energy differences $\Delta U= U_{\rm PM}(T$=0)$-U_{\rm FM}(T$=0)
 between the paramagnetic and the ferromagnetic state as a function of the 
exchange coupling $J$ for a monolayer. The negative values of $\Delta U$ at small $J$ for $n=0.6$ indicate that 
ferromagnetism is unstable.}
\label{Wkin_Wpot}
\end{figure} 

Of course in real systems, Coulomb interaction prevents charge
separation on a macroscopic scale. For the present purpose it is sufficient to discuss 
how charge transfer affects the ferromagnetic transition temperature of Kondo lattice
films. To this end we will compare results based on calculations with and
without charge transfer. Enforcing charge neutrality means we evaluated the
layer-dependent centers of gravity of the Bloch bands self-consistently
such that $n_{\alpha}=n$ for all $\alpha$. Where not otherwise stated the 
following results were obtained for charge neutral KLM films.

\subsection{Curie temperature}

\subsubsection{Whole $J$-range}

Our non-perturbative theory allows for an evaluation of KLM films at all coupling 
strengths $J$. In what follows the anisotropy constant is much smaller than the other energy 
scales in our model, $K_{2}<<W,J$. The influence of the anisotropy on the Curie temperature will be 
discussed in the last subsection in more detail. 

Figs. \ref{Tc_J_n0_2} and \ref{Tc_J_n0_6} display the ferromagnetic critical temperature as a 
function of the exchange coupling $J$ for various film thicknesses. The inset in Fig. \ref{Tc_J_n0_2} 
shows $T_{C}$ of a monolayer for a small band occupation $n=0.01$. Here one 
recognizes the typical $T_{C}\sim J^2$-behavior of the perturbational RKKY interaction 
for small $J$. Increasing the electron density induces a critical interaction $J_c$ below which 
ferromagnetism is not stable. For a given number of layers $J_c$ increases with $n$. It has been 
shown before analytically that there is a minimum critical interaction 
strength for stable FM in the KLM in infinite dimensions.\cite{CMS00}

To discuss this point further we consider the internal energy per site $U$:
\begin{equation}
U=\langle H \rangle=\langle H_{0} \rangle + \langle H_{J} \rangle = U_{\rm kin} + U_{\rm pot}\;.
\end{equation}
It can easily be calculated with the LDOS (\ref{ldos}):
\begin{equation}
  \label{Utot}
  U=\frac{1}{N_{L}}\sum_{\gamma\sigma}\int_{-\infty}^{+\infty}dEf(E)E\rho_{\gamma\sigma}(E)\;.
\end{equation}
We start from $F=U-TS$ where $F$ is the free energy and $S$ is the entropy. By comparing the difference 
of the internal energy at $T$=0 between the ferromagnetic (FM) and paramagnetic (PM) state, 
$\Delta U= U_{\rm PM}(T$=0$)-U_{\rm FM}(T$=0), we can evaluate the stability of the ferromagnetic 
state against paramagnetism. 
In Fig. \ref{Wkin_Wpot} the kinetic, potential, and total energy differences are shown 
for a monolayer at $T$=0. The magnetic stability 
for low to moderate $J$ is governed by a complex interplay between $U_{\rm kin}$ and $U_{\rm pot}$. Whereas at 
$n=0.2$ it is the potential energy which favors the (ferro)magnetic state it is both 
energies which make the ferromagnetic state unstable for higher $n$, leading to a critical interaction 
strength as in the self-consistent calculation based on the effective Heisenberg model before. The 
fact that in terms of the ground state energy there is no $J_c$ at $n=0.2$ contrary to the results 
in Fig. \ref{Tc_J_n0_2} suggests that other magnetic correlations such as non-commensurate or 
antiferromagnetic become important, reducing further the parameter range of FM stability.  
\begin{figure}[t]
\epsfig{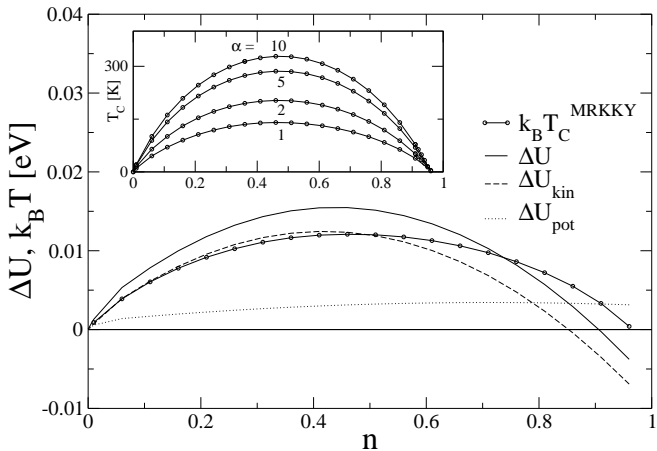}
\caption{Total, kinetic, and potential energy differences between the paramagnetic and the ferromagnetic 
state $\Delta U = U_{\rm PM}(T$=$T_C) - U_{\rm FM}(T$=0$)$ as a function of the band occupation for a 
monolayer in the strong coupling regime. The Curie temperature obtained with the MRKKY theory is also 
included for comparison. Inset: MRKKY Curie temperature for various numbers of layers. Parameters: 
$J=1~{\rm eV}, K_{2}=10^{-6}~{\rm eV}$.}
\label{DWkin_DWpot_new}
\end{figure} 

Furthermore it is remarkable that the critical interaction strength is smallest for a monolayer whereas 
films with $\alpha>1$ behave rather bulk-like as far as $J_c$ is concerned. Taking into consideration the 
energy scales of the KLM one would expect this tendency. Quite generally the physics should be governed 
by the ratio $J/W$. Hence a smaller bandwidth due to a reduced number of layers implies a smaller $J$ around 
which the intermediate coupling regime is located. In addition the reduced bulk FM region might 
indicate favored antiferromagnetic configurations which are not existent in the 2D case for mere geometrical 
reasons.

For comparison we performed calculations where charge transfer was permitted. Allowing charge 
transfer means that we used equal center of gravities of the non-interacting local density of states 
$T_{0}^{\alpha}=T_{0}$ for all layers. We did not find any noteworthy influence of charge transfer on $T_C$ 
compared to the charge neutral case as Fig. \ref{Tc_J_n0_2} demonstrates for a 5-layer film.\cite{remark3}  

\begin{figure}[t]
\epsfig{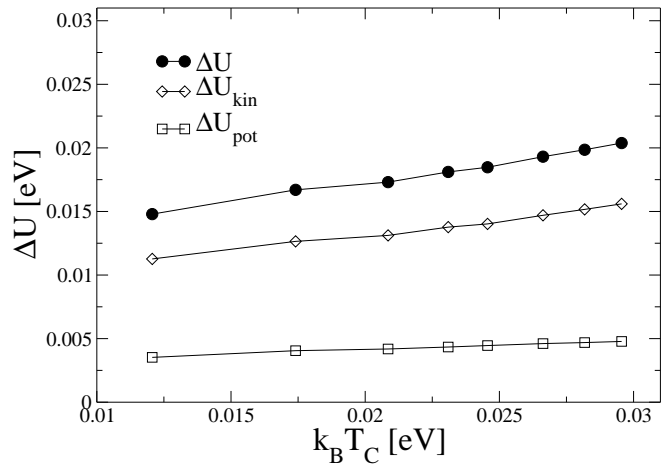}
\caption{Total, kinetic, and potential energy differences between the paramagnetic and the ferromagnetic 
state $\Delta U = U_{\rm PM}(T$=$T_C) - U_{\rm FM}(T$=0$)$ as a function of the Curie temperature in the 
strong coupling regime. The film thickness is an implicit parameter and increases from left (monolayer) 
to right ($N_L=15$). Parameters: $n=0.5, J=1~{\rm eV}, 
K_{2}=10^{-6}~{\rm eV}$.}
\label{Wkin}
\end{figure} 
\begin{figure}[t]
\epsfig{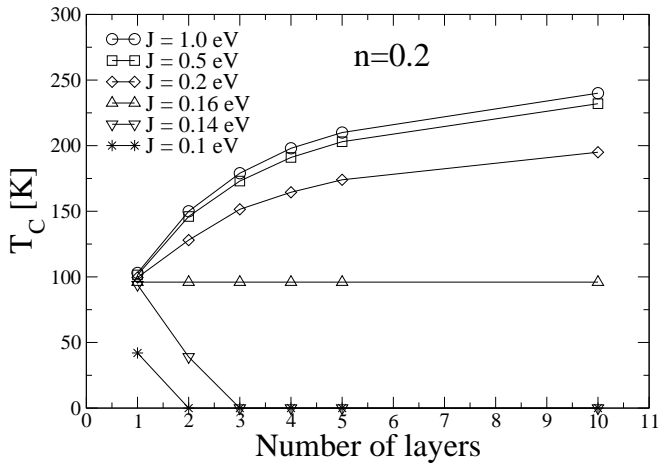}
\caption{Curie temperature as a function of the film thickness. Lines are guides to the eye. 
Parameters: $n=0.2, K_{2}=10^{-6}~{\rm eV}$.}
\label{Tc_d}
\end{figure}

\subsubsection{Double exchange regime}

In the strong coupling (double exchange) regime ($JS>>W$) one observes that $T_C$ runs into saturation 
for large enough $J$ at all electron densities and numbers of layers. Once more we will contrast the 
results obtained by the MRRKY method with energy considerations. Whereas before we were interested in the stability 
of ferromagnetism at $T$=0 we want to be more quantitative in the following comparison of the Curie temperatures. 
Hence we evaluate the internal energy of the paramagnetic state at $T_{C}$ as the energy scale for the critical 
temperature is set by the change in total energy when going from the ferromagnetic to the paramagnetic 
state $T_{C}\sim \Delta U = \Delta U(T_{C}) = U_{\rm PM}(T$=$T_C) - U_{\rm FM}(T$=0$)$. In the double exchange 
regime $T_{C}$ is essentially determined by the kinetic energy: in the local frame the electron spin is oriented 
parallel to the localized spin and the potential energy does not change much between the ferromagnetic and the 
paramagnetic state; for $J\rightarrow \infty$ it stays the same. On the other hand hopping of electrons strongly 
depends on the magnetic configuration and is somehow blocked when the localized spins are disordered. For the 
bulk KLM the relationship between $T_C$ and $\Delta U_{\rm kin}$ has already been analysed.\cite{CMS00,CMS03} 

The inset of Fig. \ref{DWkin_DWpot_new} shows the Curie temperature as a function of the band occupation $n$ for strong 
coupling. We obtain ferromagnetism for a wide range of $n$ up close to half-filling and a 
symmetry approximately around quarter-filling for all films. $T_C$ increases with the number 
of layers. The maximum band occupation for which ferromagnetism exists does not depend on the film 
thickness, indicating that the FM phase boundary in the strong coupling regime is the same in 2D and 3D.
Also in Fig. \ref{DWkin_DWpot_new} a comparison between $\Delta U$ and $T_C$ is shown. 
The variation of the critical temperature with $n$ is paralleled by a corresponding variation of 
$\Delta U_{\rm kin}$. $\Delta U_{\rm pot}$  varies only slowly with $n$. We find that, apart from a small 
difference in the maximum $n$ yielding ferromagnetism, there is even quantitatively a good agreement 
between the values of $T_C$ and $\Delta U$. 

In Fig. \ref{Wkin} we show $\Delta U_{\rm kin}$ and 
$\Delta U_{\rm pot}$ as functions of $T_C$ with the number of layers as implicit parameter. One recognizes 
a linear dependence between the ferromagnetic transition temperature and the energy change. $T_C$ is 
again dominantly determined by the difference in kinetic energy of the FM and of the PM phase. We point 
out that the almost quantitative agreement of $T_{C}^{\rm MRKKY}$ and $\Delta U$ for the monolayer is due 
to the particular choice of the anisotropy parameter and thus rather coincidental. With an increasing 
number of layers deviations between the two quantities appear. However the magnitudes remain similar.  

\begin{figure}[t]
\epsfig{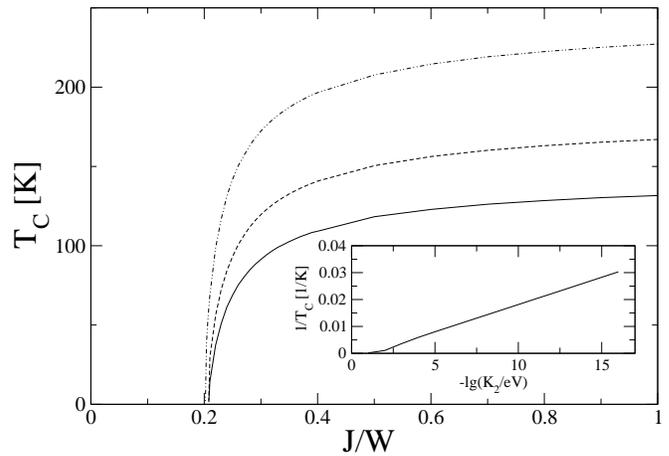}
\caption{Curie temperature of a monolayer at different anisotropy strengths for $n=0.6$ (from bottom to top: 
$K_{2}=1,~10,~100~\mu$eV). Varying the 
anisotropy has a considerable effect on $T_C$ in the strong coupling regime but only slightly changes $J_c$. Inset: 
Curie temperature as a function of $K_{2}$ for a monolayer in the saturation region ($n=0.2, J=0.2~{\rm eV}$, 
see also Fig. \ref{Tc_J_n0_2}).}
\label{Tc_J_n0_6_K2}
\end{figure}

\subsubsection{$T_C$ as a function of film thickness}

We conclude this section by presenting the critical temperature as a 
function of the film thickness. As a natural consequence of the results and the discussion just presented this 
dependence varies drastically in the different coupling regimes as demonstrated in Fig. \ref{Tc_d}. Whereas 
for large $J$ the Curie temperature decreases when reducing the number of layers, reflecting the findings 
just discussed before in Fig. \ref{Wkin}, we get a completely different picture for weak to intermediate strength 
of $J$. At $J=0.16$ the ferromagnetic transition temperature is equal for all films. Reducing further the exchange 
coupling $J$ yields a finite $T_C$ only for very thin films.

\subsubsection{Anisotropy}

In the last subsection we want to briefly discuss the effect of a single-ion anisotropy on the ferromagnetic transition 
temperature of Kondo lattice films. There exists abundant literature on scaling relations between the critical temperature 
and key film parameters as anisotropy strength, exchange coupling, and the film thickness, mostly for Ising-like 
or Heisenberg films. As we are dealing with an effective localized spin model, derived from the KLM, these results should 
also apply in our case. The exchange integrals (\ref{Jeff}) now are, however, $T$-dependent. 
In the following investigation of the functional dependence $T_C(J,K_{2})$ we restrict ourselves to a monolayer 
because in this case the role of anisotropy is most pronounced and the extension to multilayers is straightforward.

The effect of the anisotropy strength on the ferromagnetic transition temperature over the whole range of the intra-atomic 
exchange coupling $J$ is shown in Fig. \ref{Tc_J_n0_6_K2}. One can distinguish two regimes:
As can be seen from the inset (representing the saturation regime) the well known logarithmic dependence of the Curie 
temperature on the anisotropy strength $K_2$ is also valid for the Kondo lattice model. Already a very small anisotropy is sufficient 
to obtain a $T_C$ which is of the order of magnitude of Curie temperatures realized for much larger, more realistic values of $K_2$. 

Evaluating the formula for the Curie temperature 
(\ref{Tc}) by solving the corresponding $q$-integral for the predominantly contributing small wave vectors one gets
\begin{eqnarray}  
\label{Tc_scaling}
\frac{1}{T_{C}}\propto ln \left(1+ c\frac{\tilde{J}_{T=T_{C}}}{K_{2}}\right)\;.
\end{eqnarray}  
$c$ is a constant $\approx 1$ and $\tilde{J}$ is the $T-$dependent effective coupling of one localized spin to all 
others, i.e. it determines the order of magnitude of the magnon energies, which are typically much larger than 
the anisotropy energy, $\tilde{J}>>K_{2}$. An analogous formula for the Heisenberg model with constant exchange coupling is 
discussed, e.g., in Ref. \onlinecite{FJK00}. In our case the $T$-dependence of the exchange integrals (\ref{Jeff}) in the 
paramagnetic regime is evidently too weak to cause significant deviations from the logarithmic dependence, not too surprisingly because this 
temperature dependence is a mere Fermi softening effect. 
 
Whereas in the saturation region the scaling of $T_C$ with 
$K_2$ is determined by Eq. (\ref{Tc_scaling}) as just discussed, in the critical region FM becomes unstable and a higher $K_2$ only slightly 
reduces $J_c$. There $T_{C}\rightarrow 0$ and thus the effective coupling $\tilde{J}\rightarrow 0$, see Eq. 
(\ref{Tc}) where the effective coupling corresponds to the ${\hat J}$-exchange term. The critical 
region has a width of $\Delta J\approx 10^{-1}$ eV within which the effective 
coupling typically changes by $\Delta\tilde{J}\approx 10^{-3}$ eV. Thus a $\Delta K_{2}\approx 10^{-4}$ eV should 
shift $J_c$ by $\approx 10^{-2}$ eV, which is in accordance with the numerical findings in Fig. \ref{Tc_J_n0_6_K2}.

\section{Conclusions}

We have investigated ferromagnetic Kondo lattice films with a finite band occupation. Our self-consistent 
theory is made up by an electronic and a magnetic part. The one particle Green function is obtained 
within an equation of motion approach. Local moment correlation functions are calculated by an effective 
Heisenberg Hamiltonian which results from integrating out the electronic degrees of freedom. 

We presented 
results for the temperature dependent one particle exciation spectrum and discussed charge transfer between film 
layers. Charge transfer occurs in all coupling regimes and, at strong coupling, leads to a pronounced deviation 
of the surface layer electron density from the rather homogeneous band occupation of the inner layers. The 
reduced dimensionality is also responsible for a reduced magnetization at the surface. However, $T_C$ is the same for all layers. 

Curie temperatures were discussed for a variety of parameters like the exchange coupling 
$J$, the band occupation $n$, and for various number of layers. For large enough band occupation we obtain a 
critical $J_c$ below which ferromagnetism is not stable. In the strong coupling regime $T_C$ saturates and 
increases with the number of layers. We have not found any significant influence of charge transfer on the Curie 
temperature of KLM films.

In addition to the MRKKY self-consistent $T_{C}$-results energy calculations 
of the paramagnetic and ferromagnetic state were used to discuss both the critical interaction strength 
in the intermediate coupling regime and the crucial role the kinetic energy plays in the strong coupling regime.
Depending on the values of the exchange 
coupling and the band occupation the dependence of the Curie temperature on the film thickness exhibits quite a 
different behavior. We also discussed the effects of a single-ion anisotropy on the ferromagnetic transition 
temperature. Whereas it can significantly elevate $T_C$ in the strong coupling regime it hardly affects the 
critical interaction strength $J_c$. 

In this work we took the rather classical spin quantum number $S$=7/2. This would apply for rare earth systems 
as, e.g., Gd. Systems of current interest as DMS ($S$=5/2) possess a somewhat high $S$, and even for the manganites 
with $S$=3/2 classical spins are believed to be a reasonable approximation at least at low temperatures.\cite{KN06} 
Nevertheless a more profound investigation of the dependence of the magnetic properties on the spin quantum number 
at higher temperatures appears to be highly valuable.

A worthwhile extension of the present work would also be to use other lattice structures and to take
additional magnetic configurations like antiferromagnetism into account. A combined bandstructure many-body 
calculation for Gd could contribute to the much debated issues of an enhanced surface Curie temperature and 
electronic surface states in this substance. More on the model level, making the exchange and anisotropy 
parameters layer dependent or analysing the influence of non-magnetic cap layers on the magnetic stability of 
KLM films would be interesting tasks. Furthermore an application to the problem of interlayer exchange 
coupling appears to be rewarding because the temperature dependent effective exchange integrals give direct 
access to the coupling between the layers in a film system. This investigation is planned for future work.


\begin{thebibliography}{14}

\bibitem{Zen51}
C.~Zener, Phys.~Rev. {\bf 81}, 440 (1951); Phys.~Rev. {\bf 82}, 403 (1951) 
\bibitem{AH55}
P.~W.~Anderson and H.~Hasegawa, Phys.~Rev. {\bf 100}, 675 (1955)
\bibitem{Dag03} 
E.~Dagotto, Nanoscale Phase Separation and Colossal Magnetoresistance,
The Physics of Manganites and Related Compounds, Springer Series in
Solid-State Sciences {\bf 136} (2003)
\bibitem{Ram97}
A.~P.~Ramirez, J.~Phys.~C {\bf 9}, 8171 (1997)
\bibitem{Ohno99} 
H.~Ohno, J.~Magn.~Magn.~Mat. {\bf 200}, 110 (1999)
\bibitem{ZFS04} 
I.~$\breve{Z}$uti$\acute{c}$, J.~Fabian, and S.~Das~Sarma,
  Rev.~Mod.~Phys. {\bf 76}, 323 (2004)
\bibitem{Wach64} 
P.~Wachter, Helv. Phys. Acta {\bf 37}, 637 (1964)
\bibitem{SEN04}
C.~Santos, W.~Nolting, and V.~Eyert, Phys.~Rev.~B {\bf 69}, 214412 (2004)
\bibitem{ScN01}
R.~Schiller and W.~Nolting, Phys.~Rev.~Lett. {\bf 86}, 3847 (2001)
\bibitem{Ott06}
H. Ott, S. J. Heise, R. Sutarto, Z. Hu, C. F. Chang, H. H. Hsieh, H.-J. Lin, C. T. Chen, and L. H. Tjeng,
Phys. Rev. B {\bf 73}, 094407 (2006)
\bibitem{Ploog05}
S. Dhar, L. Perez, O. Brandt, A. Trampert, K. H. Ploog, J. Keller, and B. Beschoten, Phys.~Rev.~B {\bf 72}, 245203 (2005)
\bibitem{Sid06}
A. A. Sidorenko, G. Allodi, R. De Renzi, G. Balestrino, and M. Angeloni, Phys.~Rev.~B {\bf 73}, 054406 (2006)
\bibitem{Wang05}
K. Y. Wang, K. W. Edmonds, R. P. Campion, L. X. Zhao, C. T. Foxon, and B. L. Gallagher,
Phys.~Rev.~B {\bf 72}, 085201 (2005)
\bibitem{Wand98} 
M.~Farle and K.~Baberschke in: Magnetism and Electronic Correlations in Local-moment 
Systems, M.~Donath, P.~A.~Dowben, and W.~Nolting (Eds.), p. 35, World Scientific (1998)
\bibitem{SPF00}
A. B. Shick, W. E. Pickett, and C. S. Fadley, Phys.~Rev.~B {\bf 61}, R9213 (2000)
\bibitem{Nick04}
C. L. Nicklin, M. J. Everard, C. Norris, and S. L. Bennett, Phys.~Rev.~B {\bf 70}, 235413 (2004)
\bibitem{KN04}
J. Kienert and W.~Nolting, J.~Mag.~Mag.~Mat. {\bf 272-276}, e887-e888 (2004)
\bibitem{Urb94}
A.~Urbaniak-Kucharczyk, phys. stat. sol. (b) {\bf 186}, 263 (1994)
\bibitem{MN04}
W.~M\"uller and W.~Nolting, Phys.~Rev.~B {\bf 69}, 155425 (2004)
\bibitem{SN99}
R.~Schiller and W.~Nolting, Phys.~Rev.~B {\bf 60}, 462 (1999)
\bibitem{NRM97}
 W.~Nolting, S.~Rex, and S.~Mathi~Jaya, J. Phys.: Condens. Matter {\bf
   9}, 1301 (1997)
\bibitem{SN02}
C.~Santos and W.~Nolting, Phys.~Rev.~B {\bf 65}, 144419 (2002)
\bibitem{roldiss00}
R.~Schiller, Correlation Effects and Temperature Dependencies in Thin Ferromagnetic Films: 
Magnetism and Electronic Structure, Dissertation, Humboldt-Universit\"at zu Berlin (2000)
\bibitem{BMG05}
For a recent investigation of Kondo scaling in the ferromagnetic Kondo lattice see 
S. Biermann, L. de' Medici, and A. Georges, Phys. Rev. Lett. {\bf 95}, 206401 (2005)
\bibitem{Nol7} 
W.~Nolting, Grundkurs Theoretische Physik 7, Viel-Teilchen-Theorie,
Springer Berlin (2005)
\bibitem{RKKY}
A. A. Rudermann and C. Kittel, Phys. Rev. {\bf 96}, 99 (1954);
T. Kasuya, Prog. Theor. Phys. {\bf 16}, 45 (1956);
K. Yosida, Phys. Rev. {\bf 106}, 893 (1957)
\bibitem{NolQdM} 
W.~Nolting, Quantentheorie des Magnetismus, Teubner Stuttgart (1986)
\bibitem{Tya69}
S.~V.~Tyablikov, Quantentheoretische Methoden des Magnetismus, Teubner
Stuttgart (1969) 
\bibitem{Cal63}
H.~B.~Callen, Phys.~Rev. {\bf 130}, 890 (1963)
\bibitem{MW66}
N.~D.~Mermin and H.~Wagner, Phys. Rev. Lett. {\bf 17}, 1133 (1966)
\bibitem{GN01}
A. Gelfert and W. Nolting, J. Phys.: Condens. Matter {\bf 13}, R505 (2001)
\bibitem{AC64}
F.~B.~Anderson and H.~B.~Callen, Phys.~Rev. {\bf 136}, A1068 (1964)
\bibitem{SKN05}
S.~Schwieger, J. Kienert, and W.~Nolting, Phys.~Rev.~B {\bf 71}, 024428 (2005)
\bibitem{remark1}
For the Lines decoupling (M.~E.~Lines, Phys. Rev. {\bf 156}, 534 (1967)): $\gamma=\frac{9}{2S(S+1)}$
\bibitem{Nol78}
W.~Nolting, J.~Phys.~C: Solid State Phys. {\bf 11}, 1427 (1978)
\bibitem{remark2}
The fact that a spin-spin interaction of RKKY-type is long-ranged 
does not affect the argument.
\bibitem{CMS00} 
A.~Chattopadhyay, A.~J.~Millis, and S.~Das~Sarma,
  Phys. Rev. B {\bf 61}, 10738 (2000)
\bibitem{remark3}
However we point out that the addition of Coulomb interactions, which are known to be relevant e.g. for 
manganites, can lead to a significant amount of charge transfer and a variety of magnetic phases 
that occur in superlattice structures, see a recent work of C. Lin, S. Okamoto, and A.J. Millis, 
Phys. Rev. B {\bf 73}, 041104(R) (2006)
\bibitem{CMS03} 
B.~Michaelis and A.~J.~Millis,
  Phys. Rev. B {\bf 68}, 115111 (2003)
\bibitem{FJK00} 
P. Fr\"obrich, P.J. Jensen, and P.J. Kuntz,
  Eur. Phys. J. B {\bf 13}, 477 (2000)
\bibitem{KN06} 
J.~Kienert and W.~Nolting,
  Phys. Rev. B {\bf 73}, 224405 (2006)
\end{thebibliography}
\end{document}